\documentclass[prresearch,superscriptaddress]{revtex4-2}

\usepackage{adjustbox}
\usepackage{amssymb}
\usepackage{bm}
\usepackage{makecell}
\usepackage{microtype}
\usepackage{multirow}
\usepackage{paralist}
\usepackage{rotating}
\usepackage{indentfirst}
\usepackage{float}
\usepackage{mdframed}
\usepackage{float}

\usepackage{hyperref}
\usepackage[all]{xy}
\usepackage{dcolumn}
\usepackage{amsmath, amsthm, amssymb,amscd, mathrsfs, amsfonts, mathtools}
\usepackage{tikz}
\usepackage{diagbox}
\usepackage{cleveref}
\usetikzlibrary{matrix,arrows,decorations.pathmorphing}
\usetikzlibrary{backgrounds}
\usetikzlibrary{shapes.geometric}
\usepackage{circuitikz}
\usepackage{tcolorbox}
\usepackage{blkarray}
\usepackage{subcaption}
\usepackage{multirow}

\usepackage{xspace}
\usepackage{ifthen}
\newboolean{showcomments}
\setboolean{showcomments}{true}
\ifthenelse{\boolean{showcomments}}
{
}

\newcommand{\ie}{i.e.\@\xspace}

\newcommand{\etal}{et al.\@\xspace}

\newcommand{\cels}{{\em C. elegans}\@\xspace}

\def\rTo{\longrightarrow}
\def\rto{\rightarrow}

\def\<{\langle}
\def\>{\rangle}
\def\path{\rightsquigarrow}

\newcounter{para}

\newcounter{note}

\let\cal\mathcal % Fontes pour les cat�gories 
\def\b{\beta}

\def\A{\operatorname{{\textsf{A}}}}
\def\NEPb{\operatorname{NEP_\b}}
\def\sfd{\operatorname{\bf sfd}}
\def\IC{\operatorname{\bf IC}}
\def\xx{\operatorname{{\bf x}}}
\def\pp{\operatorname{{\bf p}}}
\def\kk{\operatorname{{\bf k}}}

\def\ee{\operatorname{\bf e}}
\def\ff{\operatorname{\bf f}}

\def\bc{4.2958}
\def\bs{10.7394}
\def\bfmin{1.02\b_c}
\def\bfmax{1.07\b_c}

\tikzset{every loop/.style={looseness=10, in=-130, min distance=1mm}} %% loop options
\tikzset{1simpl/.style={->,>=stealth,thick}}
\tikzset{vertex/.style = {circle, draw, color=black!60, fill=blue!20, inner sep=4}}
\tikzset{edgeto/.style={line width=.6, draw, arrows={-latex}, color=black!60 }
 }

{\endtcolorbox}

\allowdisplaybreaks

\begin{document}

\title{Brain functions emerge as thermal equilibrium states of the connectome}
\author{Elka\"ioum M.~Moutuou}
\email{elkaioum.moutuou@concordia.ca}
\affiliation{Department of Electrical and Computer Engineering, Concordia University, Montreal, QC, H3G 1M8}

\author{Habib Benali}
\email{habib.benali@concordia.ca}
\affiliation{Department of Electrical and Computer Engineering, Concordia University, Montreal, QC, H3G 1M8}

\begin{abstract}
A fundamental idea in neuroscience is that cognitive functions—such as perception, learning, memory, and locomotion—are shaped and constrained by the brain's structural organization. Despite significant progress in mapping and analyzing structural connectomes, the principles linking the brain's physical architecture to its functional capabilities remain elusive. Here, we introduce an algebraic quantum model to bridge this theoretical gap, offering new insights into the relationship between the connectome and emergent brain functions, while connecting structural data to functional predictions. Using the well-mapped \cels anatomical and extrasynaptic connectomes, we demonstrate that brain functions, defined as functional networks of a neural system, emerge as thermal equilibrium states of an algebraic quantum system derived from the graph algebra of the underlying directed multigraph. Specifically, these equilibrium states, characterized by the Kubo-Martin-Schwinger (KMS) formalism, reveal how individual neurons contribute to functional network formation.  Our model illuminates the structure-function relationship in neural circuits through two key features: (1) a functional connectome that delineates topologically driven neuronal interactions and (2) an Integration Capacity (IC) index that quantifies how effectively neurons coordinate and modulate diverse information flows. Together, these features provide a statistical and mechanistic account of information flow and reveal how the network topology of the connectome predicts cognition and complex behaviors. 
\end{abstract}

\maketitle

\section{Introduction}
A core tenet of {\em connectomics} in neuroscience is that brain functions emerge from contextual collective properties of its structural substrates~\cite{hartwell1999molecular,Bargmann2013,emmons2012mood}. It emphasizes the importance of the brain's structural network, or the {\em connectome}~\cite{sporns2005,sporns2011connectome,DeFelipe2010}, in determining how information is processed, integrated, and transmitted within such a complex system. Based on this premise, exploring the relationships between the structure of the brain's wiring systems and  the flow of information across synaptic circuitries has the potential to provide deeper insights into how neuronal assemblies~\cite{bullmore2012,Pulvermueller2014} underpin cognition, behavior, neurological and psychiatric states~\cite{Doyon2005,Yu2021}. This perspective has fueled intense research over the past few decades, aimed at: (i) producing comprehensive maps that detail the anatomical connections between neural units at various levels---from macroscale networks of brain regions and white matter pathways using neuroimaging techniques~\cite{Bellec2006,Desikan2006} to microscale synaptic circuits using high-resolution imaging technologies such as electron microscopy~\cite{white1986structure,cook2019connectome,Witvliet2021,winding2023insect,Veraszto2024}---and (ii) developing quantitative models for understanding how these complex architectures contribute to brain functions and dysfunctions~\cite{friston1993functional,Messe2014,Bettinardi2017,park2013struct,Sporns2018}. 

Large-scale connectomics data have been generated across a wide array of species ---such as \cels~\cite{white1986structure}, {\em Drosophila}~\cite{winding2023insect}, {\em Platynereis dumerilii}~\cite{Veraszto2024}, mice~\cite{Abbott2020} and humans~\cite{ShapsonCoe2024} ---and interdisciplinary research, particularly the application of graph theory and network science, has been pivotal in analyzing these datasets, modeling the interactions and information flow in neural circuits, and detecting connectivity patterns crucial for understanding the functional states of the nervous system~\cite{bullmore2012,morone2019}.

 With its 302 neurons and approximately 13,000 chemical and electrical synapses, the \cels synaptic connectome is the most complete of these endeavors~\cite{white1986structure,Varshney2011,cook2019connectome}. It has served as a prototypical nervous system for building sophisticated models to uncover specific neural circuits underlying complex behaviors and predict brain dysfunctions based on structural patterns~\cite{Yan2017,morone2019} . 

While these efforts have provided new perspectives on the structural mechanisms underlying cognition, a comprehensive theoretical framework that explains how the brain's physical architecture shapes and predicts its function is still lacking. Here, we introduce such a framework using \cels~\cite{Varshney2011,cook2019connectome} as a model connectome and applying the AQM methods for information flow that we recently developed~\cite{Moutuou2024a}. Specifically, we demonstrate how the topology of the structural connectome predicts its functional networks as KMS states of the algebraic quantum system defined by the graph algebra of the underlying directed multigraph, where neurons serve as nodes and synapses and gap junctions as directed edges.

 The concept of KMS states on complex networks, originating from graph C*-algebras~\cite{cuntz1980markov,raeburn2005graph} and the algebraic extension of quantum mechanics~\cite{haag1967quantum,hugenholtz1972}, has been shown in~\cite{Moutuou2024a} to provide statistical descriptors of the dynamic patterns of neuronal interactions. We investigate the functional implications of these descriptors and extract two fundamental features that characterize the functional organization of the \cels nervous system: (i) the {\em purely topological functional connectome} (PTFC) which maps the entire atlas of functional connections driven by the anatomical network, and (ii)  an {\em integration capacity} (IC) index that measures the degree to which neurons can modulate multiple independent flows. 

Ultimately, the algebraic quantum model presented here offers a theoretical framework for decoding the structure-function relationship in terms of physical processes, revealing neuronal assemblies that mediate sensorimotor processing and trigger complex behaviors such as thermotaxis~\cite{mori2007,kuhara2011}, mechanosensation~\cite{kaplan1993}, and locomotion~\cite{white1986structure,chalfie1985neural,Bono2005}.

The remainder of the paper is structured as follows. Section~\ref{sec:theory} introduces the theoretical framework and key derivations; Section~\ref{sec:results} presents the main results on the structure-function relationship in \cels; and Section~\ref{sec:experimental} details the experimental validations. We conclude with a discussion of functional implications and broader perspectives in Section~\ref{sec:discussion}. 
% A summary of the main notations used throughout is provided at the end of Section~\ref{sec:methods}.

\section{Theory}\label{sec:theory}

\subsection{The connectome as a directed multigraph}
We model the structure of the adult hermaphrodite \cels somatic connectome as a directed graph $G=(V, E)$ with parallel edges and self-loops, where the node set $V$ represents $N=280$ individual neurons, and the edge set $E$ consists of $12071$ unique chemical and electrical synapses (gap junctions), constructed by merging the old dataset~\cite{wormatlas2023,white1986structure, Varshney2011} with the newly revised one~\cite{cook2019connectome} (see Methods for details and Supplementary Data 1~\cite{MoutuouData2025} for the resulting edgelist). The {\em source} $s(e)$ and {\em range} $r(e)$ of an edge are respectively the pre- and post-synaptic neurons of the synapse. 
Since gap junctions allow bidirectional transmission between neurons, they are represented by reciprocal edges. Our main focus being on the functional interactions among neurons through anatomical pathways, we consider both chemical and electrical synapses within a unified wiring system, since both transmission modalities are known to maintain close functional interactions~\cite{pereda2014}.  We let $\A$ be the {\em adjacency matrix} of $G$. Specifically, since $G$ is a directed muligraph, $\A$ is an integer-weighted non-symmetric matrix, where for two neurons $u$ and $v$, the entry $\A_{uv}$ counts the number of chemical and electrical connections from $v$ to $u$. We use bold letters $\ee, \ff$, etc. for directed walks; \ie, sequences $e_1\cdots e_n$ (with possible recurrences) of edges such that $s(e_i)=r(e_{i+1})$. The length $|\ee|$ of such a walk $\ee$ is the number $n$ of edges composing it, and we set $s(\ee)=s(e_n)$ and $r(\ee)=r(e_1)$ to be its source and range, respectively. $\ee$ is a {\em cycle} if $s(\ee)=r(\ee)$. Since $G$ contains directed cycles and self-loops , finite length walks form an infinite set that we denoted by $E^*$.  Indeed, 44 neurons have synapses onto themselves (such connections are referred to as  '{\em autapses}' in the literature~\cite{bekkers2003autapse, tamas1997autapse}).

\subsection{Neural emittance}

\paragraph{Neural emittance profiles from KMS states.}
Given $\b>0$, the {\em emittance volume} of $v$ at inverse temperature $\b$ is defined as
\begin{eqnarray}\label{eqn:emittance}
Z^\b_v = \sum_{\ee \in E^*, s(\ee)=v}e^{-\b |\ee|},
\end{eqnarray}
which is an infinite series that encodes the volume of information flow downstream of $v$ when the system functions at inverse temperature $\b$; \ie, when the functioning of each connection is affected by a factor of $e^{-\b}$. We shall note that similar formulation was used for simple undirected graphs by Katz~\cite{katz1953} to measure node status in sociometry, and Estrada and Hatano in~\cite{estrada2008communicability} to study communicability in undirected complex networks. Observe that $Z^\b_v\ge 1$, reflecting the fact that a node theoretically emits onto itself through the trivial path of length $0$. In particular, if $v$ is not pre-synaptic to any neuron, its emittance volume is $1$. Furthermore, by an elementary result from functional analysis, there is a critical value $\b_c=\log r$, where $r$ is the spectral radius of $G$, such that $Z^\b$ converges for $\b>\b_c$ and 
\begin{eqnarray}\label{eq:neural-emittance}
Z^\b_v = \sum_{n=0}^\infty\sum_{u\in V} e^{-\b n}(\A^n)_{uv}= \sum_{u\in V}(1-e^{-\b}\A)^{-1}_{uv},
\end{eqnarray}
for all $\b > \b_c$. We can then use the emittance volume to quantify the following quantity
\begin{eqnarray}\label{eq:x-v-beta}
	\xx^{v|\b}_u = \frac{1}{Z^\b_v}(1-e^{-\b}\A)^{-1}_{uv}, 
\end{eqnarray} 
which, for $\b>\b_c$, we refer to as the {\em neural emittance} of $v$ to $u$ at inverse temperature $\b>\b_c$. Noticing that  $(1-e^{-\b}\A)^{-1}_{uv}=\sum_{\ee\in E^*, s(\ee)=v, r(\ee)=u}e^{-\b |\ee|}$ for $\b>\b_c$, the neural emittance of $v$ quantifies the probability for neuron $u$ to receive signals from $v$, providing a more natural and physical formulation of statistical propagation~\cite{Ghavasieh2024}.  In particular, the {\em neural self-emittance} $\xx^{v|\b}_v$ measures the probability of feedback signals onto neuron $v$ at inverse temperature $\b$. The {\em neural $\b$--emittance profile ($\NEPb$)}  of $v$ is the vector 
%\begin{eqnarray}
%	\xx^{v|\b} = (\xx^{v|\b}_u)_{u\in V},
%\end{eqnarray}
$\xx^{v|\b} = (\xx^{v|\b}_u)_{u\in V}$ encoding how information from $v$ is propagated through the structural network. That is, $\xx^{v|\b}$ quantifies how $v$ {\em functionally} connects to any other neuron when the system is at inverse temperature $\b$. Specifically, the non-zero components of $\xx^{v|\b}$ represent potential functional links of $v$ onto the corresponding neurons. 

We want to emphasize that these vectors originate from the AQM methods~\cite{hugenholtz1972} and the novel framework we developed for flow pathways in directed networks~\cite{Moutuou2024a}. Specifically, as we have shown in~\cite{Moutuou2024a}, information flow within the directed multigraph $G$ forms a system with infinitely many degrees of freedom that is best modeled by the graph C*-algebra of $G$~\cite{raeburn2005graph}. Furthermore, this graph $C^*$--algebra carries a natural dynamics that turns it into an algebraic quantum system $\cal O_G$ whose KMS states at inverse temperature $\b$---termed as $\b$--KMS states---arise from statistical superpositions of the distributions $\xx^{v|\b}$. Namely, a $\b$--KMS state of the system $\cal O_G$ is associated to a convex combination
\begin{eqnarray}\label{eq:mixed}
	\xx = \sum_v \pp_v\xx^{v|\b},
\end{eqnarray}
where $(\pp_v)_v$ is a distribution of relative quantities on neurons such as topological measures (out- or in-degree distributions), gene expression level of neurons, etc. (see~\cite{Moutuou2024a} for theoretical details). It follows that the $\NEPb$ vectors $\xx^{v|\b}$ can be interpreted as the pure $\b$--KMS states of the connectome at inverse temperature $\b$, while mixed $\b$--KMS states are expressed via Eq.~\eqref{eq:mixed}. 

To get a physical intuition of the notions of "pure" and "mixed" $\b$-KMS states, think of the pure state $\xx^{v|\b}$ as representing a state of the system where all the information flow originates from $v$, and the value $\xx^{v|\b}_u$ quantifies the probability that this flow reaches $u$. In contrast, a mixed state $\xx$ of Eq.~\eqref{eq:mixed} describe a statistical ensemble where where information flow may originates from any neuron $v$ with probabily $\pp_v$. In this case, the value $\xx_u$ corresponds to the expected information flow received by neuron $u$ in this state under the probability distribution $(\pp_v)_{v\in V}$.

One can illustrate these states by representing the vector $\xx^{v|\b}$ as a simple weighted directed star with source node $v$, where for $u\neq v$, $(v,u)$ is a directed edge of weight $\bar{\xx}^{v|\b} _u= \xx^{v|\b}_u/ \sum_{w\neq v}\xx^{v|\b}_w$ if $\xx^{v|\b}_u>0$, thus obtaining a downstream connectivity network of the neuron consisting of potential functional connections, which we refer to as its {\em emittance network}.
\medskip 

\paragraph{Structure-function divergence. }
 How and to what degree do the emittance networks of a neuron 'differ' from its structural connectivity? In order to address this question, let us consider the structural connectivity state of a neuron $v$ as the distribution $(\kk^v_u)_u$ encoding the post-synaptic density of $v$; \ie, $\kk^v_u = \A_{uv}/k^{out}_v$ where $k^{out}_v=\sum_w\A_{wv}$ is the number of chemical and electrical outgoing connections from $v$. Then, using Uhlmann's {\em transition probability} between quantum states~\cite{uhlmann1976, jozsa1994fidelity}, we define the {\em structure-function divergence} ({\bf sfd}) of a neuron $v$ as 
\begin{eqnarray}\label{eq:sfd}
\sfd(v, \b) = 1 - \left(\sum \left(\kk^v_u\bar{\xx}^{v,\b}_u\right)^{1/2}\right)^2.
\end{eqnarray}
Specifically, $\sfd$ measures how the two networks representing the $\NEPb$ diverges from the anatomical wiring of the neuron in terms of both the number and intensity of their respective connections. 

Now $\sfd(v,\b)\sim 0$ if and only if the two distributions $\kk^v \sim \xx^{v|\b}$, which would mean that when the system is in the functional state $\xx^{v|\b}$, $v$ can send signals only to its direct neighbors at the same probabilities as in its structural connectivity state. We refer to the value of $\b$ that minimizes $\max_v\sfd(v,\b)$ as the {\em structural (inverse) temperature} of the connectome, and will be denoted by $\b_s$.

\medskip

\paragraph{Functional temperatures. } On the other hand, since every $\b>\b_c$ generates emittance networks and at values close to $\b_s$ these networks do not considerably differ from anatomy, which $\b$ values should we consider functionally relevant? To address this question, we examine the {\em mean receptance} at inverse temperature $\b$, defined as the average probability for a neuron to receive signals from the other neurons, and we compare it to self-emittance. 

Specifically, for a fixed neuron $u$, consider the mixed $\b$--KMS state $\xx(u,\b)$ given by Eq.~\eqref{eq:mixed} where $\pp = (\pp_v)$ is the uniform distribution on the neuron subset $\{v \in V; \ v\neq u\}$; namely, $\pp_v = 1/(N-1)$ for $v\neq u$, and $\pp_u=0$. Then, $\xx(u,\b) = \frac{1}{N-1}\sum_{v\neq u}\xx^{v|\b}$, and the expected flow received by $u$ in this state is the value $\xx(u,\b)_u = \frac{1}{N-1}\sum_{v\neq u}\xx^{v|\b}_u$, which precisely measures the average information flow received by $u$ from all other neurons. Now, the {\em mean receptance} is obtained by taking the average $\<\xx(u,\b)_u\> = \frac{1}{N}\sum_u \xx(u,\b)$, which is given by

 \begin{eqnarray}
 \frac{\sum_v\sum_{u\neq v}\xx^{v|\b}_u}{N(N-1)}.
 \end{eqnarray}
 We then consider $\b$ as a functional inverse temperature if the mean receptance is $> 0.5$, meaning that at such a $\b$, neurons receives more signals from the other neurons than from themselves through self-emittance. 

\subsection{Dependence on network topology}
It is important to note that emittance networks result from global measures of macroscopic systems consisting of (possibly infinitely) many pathways of various depths, but they do not provide precise information about the extent to which particular synaptic circuitries are relevant to the emergence of these functional networks. More precisely, irrespective of its value, the weight of a connection in the emittance network does not, by itself, allow to know whether the occurrence of this connection is determined by the particular global topology of the anatomical network; \ie, the physical wiring of the connectome.  For example, suppose that at a certain value of $\b$ the neural emittance of $v$ onto $u$ coincides with an existing direct anatomical link. How can we know whether this particular physical wiring from $v$ to $u$ is 'essential' for a this emittance to emerge at that value of $\b$? And more generally, how dependent are the $\NEPb$ on the network topology? To address these questions, we systematically compared the weights in the emittance networks of a given neuron to those that one would get if the network topology was randomly perturbed. Specifically, we measured the likelihood for a neural emittance obtained at a fixed value of $\b$ to emerge with equal or larger weight at the same inverse temperature if the synaptic network was randomly rewired (see Methods). Based on their statistical significance, one can then distinguish the emittance connections that are determined by the network topology ($p < .05$) from those that are probably not. In a sense, the former represent the potential functional connections predicted by the network structure of the connectome; we then refer to them as {\em purely topological functional connections} (PTFC) of the neuron.

\subsection{Purely topological functional connectomes from mixed states.} 
Now notice that mixed functional states given by Eq.~\eqref{eq:mixed} can also be expressed as a matrix product via
\begin{eqnarray}\label{eq:mixed-state}
	\xx = [\xx^{\bullet|\b}]\pp,
\end{eqnarray}  
where $ [\xx^{\bullet|\b}]$ is the $N\times N$-matrix whose columns are the $\NEPb$ vectors $\xx^{v|\b}$ and $\pp=(\pp_v)_v$ is a distribution over neurons. Since each column of $ [\xx^{\bullet|\b}]$ measures potential functional connections from a neuron to the other neurons, the whole matrix measures potential functional connections between each pair of neurons in the connectome. Thus, a mixed $\b$--KMS state $\xx$ encodes 1) the potential functional connections between each neuron pair at inverse temperature $\b$, and 2) the expected value of the upstream functional connectivity onto each neuron given the distribution $\pp$ on the neurons. It follows that for each $\b$ value,  $ [\xx^{\bullet|\b}]$ is a connectivity matrix encoding the substrates underlying all functional states of the connectome. 

Moreover, using the notion of PTFCs introduced in the previous paragraph, we get a {\em purely topological functional connectivity matrix} at inverse temperature $\b$ by restricting $ [\xx^{\bullet|\b}]$ to its statistically significant values. The resulting matrix is the adjacency matrix of a weighted directed network $\cal F^\b$ whose edges the PTFCs at the given $\b$ value, and for this reason we call it the '{\em purely topological functional connectome}' at inverse temperature $\b$. Specifically, $\cal F^\b$ represents potential functional connections predicted by the topology of the anatomical network. 

\subsection{Integration capacity}
Let us now consider a particular mixed KMS state defined as follows. Letting $\pp$ in~\eqref{eq:mixed-state} be the uniform distribution $\pp_v=1/N$ on $V$, the resulting mixed state $\langle\xx^{v|\b}\rangle=\frac{1}{N}\sum \xx^{v|\b}$ is the mean $\NEPb$, and the probability of a neuron $u$ to receive signals in this state is the average neural $\b$--emittance to $u$ at inverse.

The mixed state $\langle\xx^{v|\b}\rangle$ encodes the average information pathways upstream of every neuron~\cite{Moutuou2024a}, which might involve feedback signal pathways. Thus, comparing two neurons according to the weights of their incoming functional connections in this state can be biased if, for example, most incoming paths to one neuron consists of feedback loops, while the other one mostly receives pathways that do not communicate to each other through cycles. In order to quantify the extent to which a neuron $u$ 'integrates' independent pathways, we introduce the following measure that we term as {\em integration capacity} $0\le \IC\le 1$ at inverse temperature $\b$:

\begin{eqnarray}
	\IC_u(\b) = \frac{1}{N-1}\left(N\langle\xx^{v|\b}\rangle_u - \xx^{u|\b}_u\right).
\end{eqnarray} 
Specifically, it is the expected relative weight of all possible PTFCs  onto the neuron, excluding its neural self-emittance at inverse temperature $\b$. A high $\IC$ value at a given $\b$ means a high likelihood of receiving any information flowing through the network, reflecting two topological features of the neuron: 1) it is the target of a relatively large number of independent flow pathways of certain lengths; in other words, a high capacity for the neuron to integrate information flow coming from many different neurons that do not belong to the same cycles; in particular, the $\IC$ of a highly connected neuron with nearly no closed paths will be close to 1, whereas an equally connected neuron whose almost all incoming paths are cycles will have an $\IC$ close to $0$; and 2) it has substantial redundancies among its incoming pathways; thus, the neuron is well-situated in the connectome to receive inputs from multiple neurons via a number of parallel pathways of certain range. 

In one sense, neurons whose $\IC$ is high at small temperature values but low at larger ones are structurally positioned to  mediate relatively short-range communications, making them '{\em local integrators}'~\cite{park2013struct}, whereas those with high $\IC$ at higher temperatures are '{\em global integrators}' capable of mediating multiple long-range incoming communications.

%%%% RESULTS %%%%
\section{Results}\label{sec:results}

\begin{figure*}[!ht]
	\centering
	\includegraphics[width=.8\textwidth]{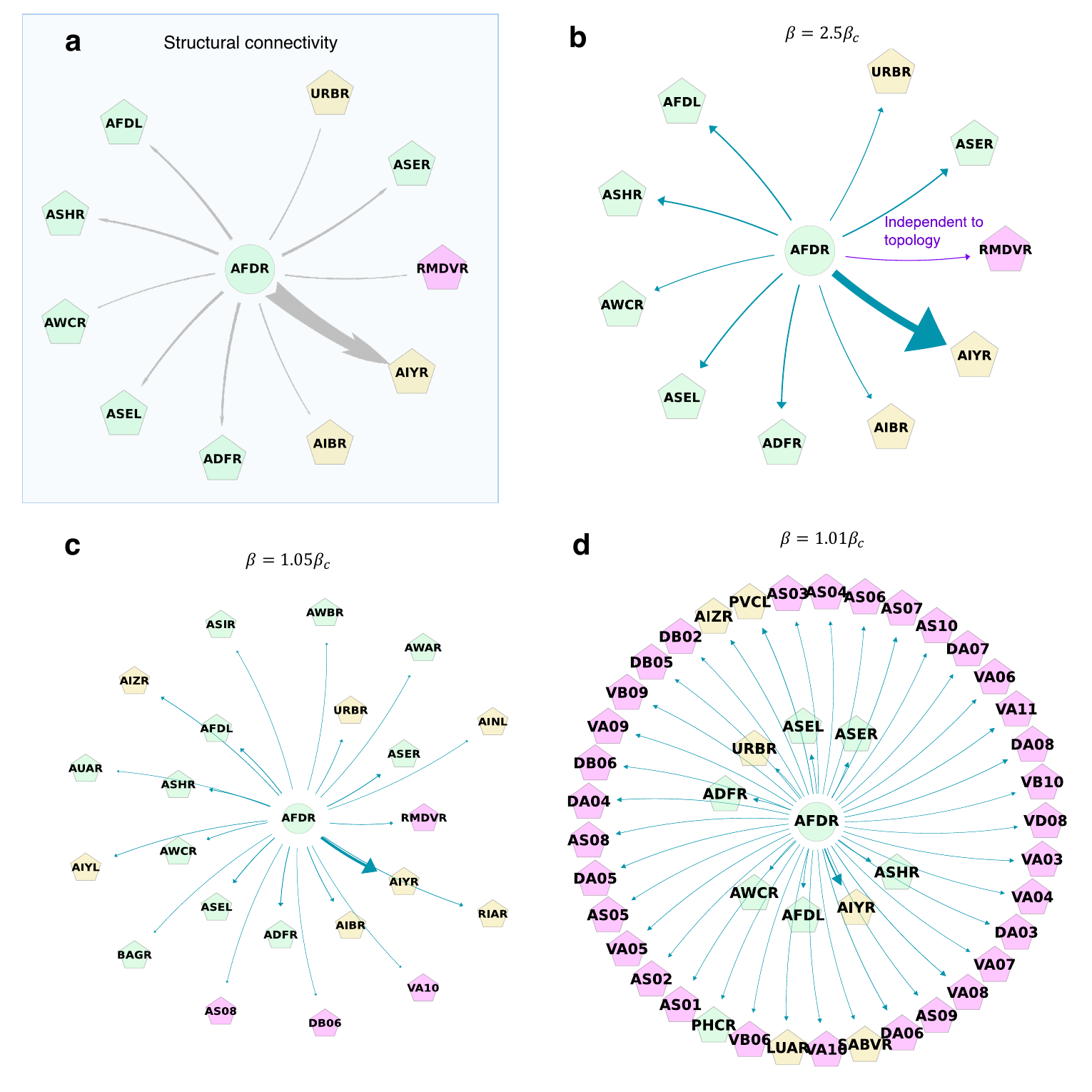}
	\caption{{\bf Connectivity states of AFDR.} Schematic visualizations of different types of connectivity states AFDR: {\em structural connectivity} (gray arrows), and the emittance networks (green and purple arrows) obtained from  {\em neural $\b$-emittance profiles} ($\NEPb$) at different values of inverse temperature  $\b>\b_c$, where $\b_c$ is the critical inverse temperature. {\bf a}: the structural connectivity of AFDR; the weights of the connections are the components of the structural connectivity vector $\kk^{AFDR}$. {\bf b}:  At $\b_s=2.5\b_c$, the emittance network coincides with the structural connectivity of AFDR. However, the connection to RMDVR (represented by the purple arrow) is not statistically significant  ($p\sim 0.2$), and therefore not a purely topological functional connection (PTFC). In other words, when the system is at this $\b_s$--KMS state, AFDR can functionally communicates only with its direct neighbors at the same probability distribution as its structural connectivity state. {\bf c}: At a relatively higher temperature $1/\b$, long-range communications emerge (\ie neurons that are not post-synaptic to AFDR, here positioned at the external concentric circle) while the neuron maintains neural emittances onto its direct neighbors (neurons positioned at the internal circle), albeit with lower intensity.  {\bf d}:  At a very high temperature $1/\b$ with $\b$ close to the critical value, more longer range connections emerge, while some short range ones are either no longer present or not statistically significant. In both {\bf c} and {\bf d}, only the PTFCs ($p< 0.05$) are considered. Greens are sensory neurons, yellows are interneurons, and pinks are motor neurons.}
	\label{fig:NEP}
\end{figure*}

\subsection{Emittance networks}
To illustrate our model, we reprent in Figs.\ref{fig:NEP}b--d the emittance networks of the thermosensory neuron $v={\rm AFDR}$ at different values of $\b$, illustrating how they evolve as the (inverse) temperature varies. We see that increasing the temperature allows the emergence of long-range emittance connections which are facilitated by large amounts of parallel pathways of different lengths. Indeed, it is an immediate consequence of formula~\eqref{eq:x-v-beta} that parallel paths and redundancies are critical to the emergence of strong neural emittances, and the lengths of pathways and the degree of redundancies that significantly contribute to their weights depend on temperature. More precisely, lower temperatures (\ie, higher values of $\b$) favor parallel direct anatomical connections and short paths. Whereas, higher temperatures enhance the contributions of redundant long pathways of different lengths while lessening the importance of direct anatomical connections in the emittance weights. For example, AFDR has 13 synaptic connections and 52 anatomical paths of length 2 onto the interneuron AIYR, contributing greatly to the important weight of its neural emittance onto the latter at low temperature (Fig.\ref{fig:NEP}b). And at higher temperature values, while the neural emittance of AFRD $\rTo$ AIYR decreases, other connections are 'established' with neurons onto which AFDR can only be connected through thousands or millions of parallel anatomical paths of length $\ge 2$. For instance, there are 117 paths of length 2 and about 4 millions of length between 3 and 5 linking AFDR to RIAR, contributing to the neural emittance onto this neuron  at $\b=1.05\b_c$ (Fig.\ref{fig:NEP}c), with $\b_c = \bc$. Similarly, the neural emittance to the ventral cord motor neuron VA11 that appears at $\b=1.01\b_c$ (Fig.\ref{fig:NEP}d) is largely due to the more than its 230 million incoming paths of length $\ge 5$ from AFDR.

\subsection{Structure vs function}

\paragraph{Computing the $\sfd$. }
We found that $\sfd(v, \b) \sim 0$ at $\b=2.5\b_c$ for each of the 280 neurons. Therefore, we have $\b_s \approx \bs$. In particular, going back to the example of AFDR, its neural $\b_s$--emittance profile has approximately the same values as its structural connectivity state. For instance, the structural connection from AFDR to the motor neuron RMDVR has weight $\kk^{AFDR}_{RMDVR}=0.037037$, and its neural emittance onto RMDVR at $\b_s$ is $\xx^{{AFDR}|\b_s}_{RMDVR}\approx 0.037032$ (see~\Cref{tab:AFDR-RMDVR}). Namely, the emittance network of AFDR at inverse temperature $\b_s$ perfectly maps over its anatomical connectivity network.

\medskip 

\paragraph{Diversity in structure-function divergence. }
More generally, the $\sfd$ measure provides the ratio of non-overlapping links in both functional states. For instance, $\sfd(AS08,\b)= 0.125$ for $\b= 1.7\b_c$ means that, at this inverse temperature, the emittance network of the motor neuron AS08 deviates from anatomy by 12.5\%. 

Analyzing $\sfd$ variation of all \cels neurons with respect to inverse temperature, we found that the emittance networks of some neurons, such as the motor neurons DD02, AVAL/R, AS04, remain close to their structural connectivity over long intervals, whereas those of other neurons, such as AS08, the polymodal sensory neurons PVDL/R and FLPL/R, and the interneurons LUAL/R, diverge rapidly from structure after certain low temperature values. For example, at $\b=7.30$, AVAL and AVAR emittance networks deviate from structure by only 1.3\%, while PVDL/R deviate by $\sim 9\%$ and AS08 deviates by 12.5\% as mentioned earlier.

\medskip
 
\paragraph{The \emph{c. elegans} functional temperature interval. } 
 By computation we found that the mean receptance of \cels is $~0.5$ at $\b_{\rm o}=1.07\b_c$ and $>0.5$ on the interval $I_f=(\b_c, \b_{\rm o})$. For instance, Fig.~\ref{fig:NEP}c and Fig.~\ref{fig:NEP}d represent the emittance networks of AFDR at two different functional inverse temperatures ($\b = 4.51$ and $\b = 4.34$, respectively). 

\medskip 

\paragraph{Purely topological functional connections. }
Analyzing the emittance network of AFDR at the structural inverse temperature $\b_s$ shown in Fig.~\ref{fig:NEP}b , we found that the emittance onto RMDVR (purple arrow) is not statistically significant ($p\sim .2$, see~\Cref{tab:AFDR-RMDVR}), therefore not a PTFC. On the other hand, noting that the anatomical connection from AFDR to RMDVR is defined by a gap-junction, hence a reciprocal connection, we investigate the structural connectivity of RMDVR along with its emittance network (\Cref{tab:AFDR-RMDVR}) at inverse temperature $\b_s$ in order to check whether there is any directional preference on the functioning of this gap-junction. Indeed, we found that the neural emittance of RMDVR onto AFDR is a PTFC ($p=.0056$, see~\Cref{tab:AFDR-RMDVR}). 

% TABLE : AFDR vs RMDVR
\begin{table}[!h]
	\centering
	\caption{{\bf Neural emittances of AFDR and RMDVR at $\b_s$}. }
	\includegraphics[width=.5\textwidth]{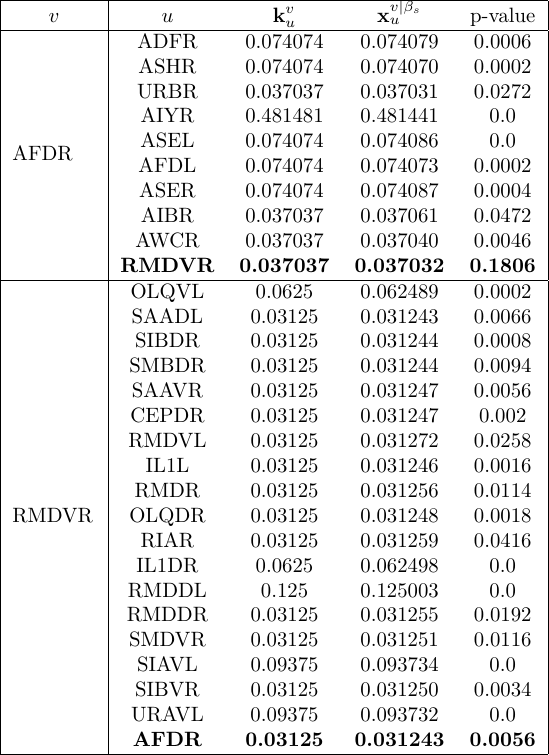}

	\label{tab:AFDR-RMDVR}
\end{table}
Note that (in)dependence of a neural emittance connection on topology might vary with (inverse) temperature change. For instance, a PTFC from AFDR to RMDVR  eventually emerges significantly ($p=.015$) at a functional temperature $\b_f=1.05\b_c$ (Fig.~\ref{fig:NEP}c), albeit with lower weight. This apparent discrepancy is a consequence of the fact that the presence of a large volume of redundant anatomical short paths between two neurons increases the likelihood of these neurons to remain functionally connected at low temperature if the system is randomly rewired, and at higher temperature, redundant long-range pathways increases the likelihood of the emittance networks to be less altered by structural rewiring. For instance, there are $1464$ anatomical paths of length $\le 3$ all connecting AFDR onto RMDVR, while longer paths between these neurons rely heavily on the existing autapses on RMDVR and other interconnecting neurons such as AFDL, RIAR, IL1DR, etc., forming an infinite number of feedback loops that are very likely destroyed after random reconfiguration of the network.  

%%%%%% RID %%%%
\begin{figure}[!h]
	\centering
	\includegraphics[width=.42\textwidth]{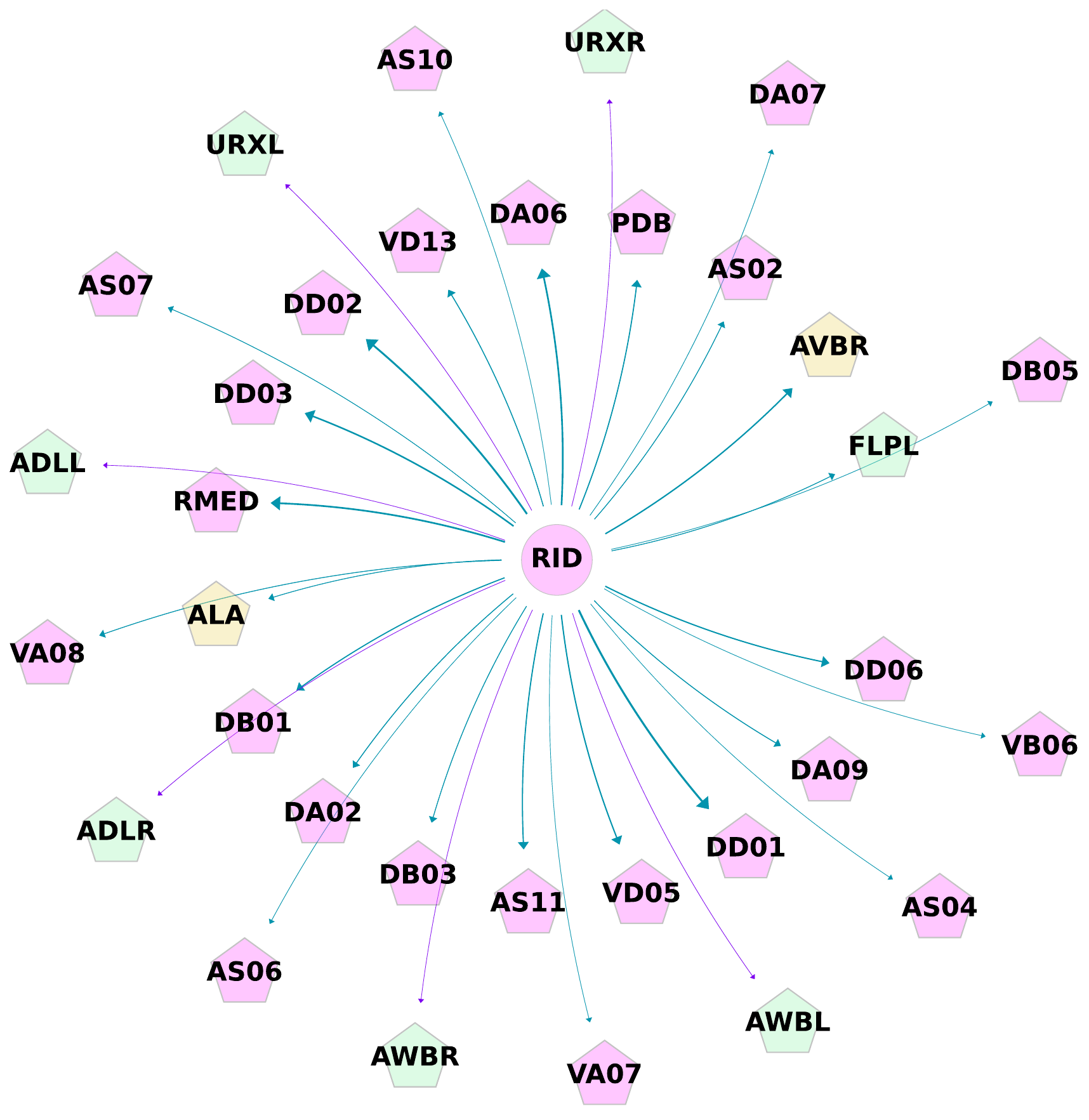}
	\caption{{\bf Emittance network of RID.} The pink arrows are connections that are not determined by the network topology, that is, the particular anatomical wiring of the neuron post-synaptic connections, therefore not PTFCs. Green arrows are emittance connections that are topology-dependent, therefore PTFCs. These connections coincide with the functional connections from RID to the neuron classes URX and ADL that have been identified by Randi \etal to be driven by extrasynaptic signaling.}
	\label{fig:RID}
\end{figure}

\subsection{Putative extrasynaptic signaling pathways}

A biological implication of emittance connections that are not PTFCs is that the involved neurons do not necessarily rely on the fixed global anatomy to functionally interact with its target neurons, and additional 'structural' connections might be needed for their functional connectivity to be predicted from anatomy. This is consistent with results from multiple studies that have demonstrated the presence of extrasynaptic signaling between neurons~\cite{Barrios2012,Lim2016,bentley2016multilayer,ripoll2022,randi2023celegans}. To test this interpretation, we investigated the PTFCs of the motor neuron and interneuron RID which has recently been identified by Randi \etal~\cite{randi2023celegans} to generate functional connections that were not predicted from anatomy. 

Specifically, we computed the $\NEPb$ of RID at a functional inverse temperature $\b_f = 1.05\b_c = 4.51$ and its emittance network is represented in~\Cref{fig:RID}.  We found that the emittance connections of RID onto ADLL/R and URXL/R are statistically non-significant ($p=1$), and its emittance connection onto AWBL/R are very weak ($\le 10^{-4}$) and statistically non-significant with $p=1$. 

On the other hand, considering an extrasynaptic chemical signaling as an actual directed physical connection between neurons, we separately added to the graph one edge from RID to URXL and one edge from RID to ADLR.  And by recomputing the $\NEPb$ of RID within the updated directed multigraphs, we found that  (see Supplementary Data 5 and 6~\cite{MoutuouData2025}) the new neural emittances from RID to URXL and ADLR are PTFCs ($p=0.0082$ and $p=0.0086$, respectively). 

These observations theoretically support the results from~\cite{randi2023celegans} that detected functional connections RID $\rto$ ADLR and RID $\rto$ URXL that are not predicted by the anatomy and that partly rely on chemical extrasynaptic transmissions.

\medskip 

%%%%% Pure functional connectome
\begin{figure*}[!h]
	\centering
	\includegraphics[width=\textwidth]{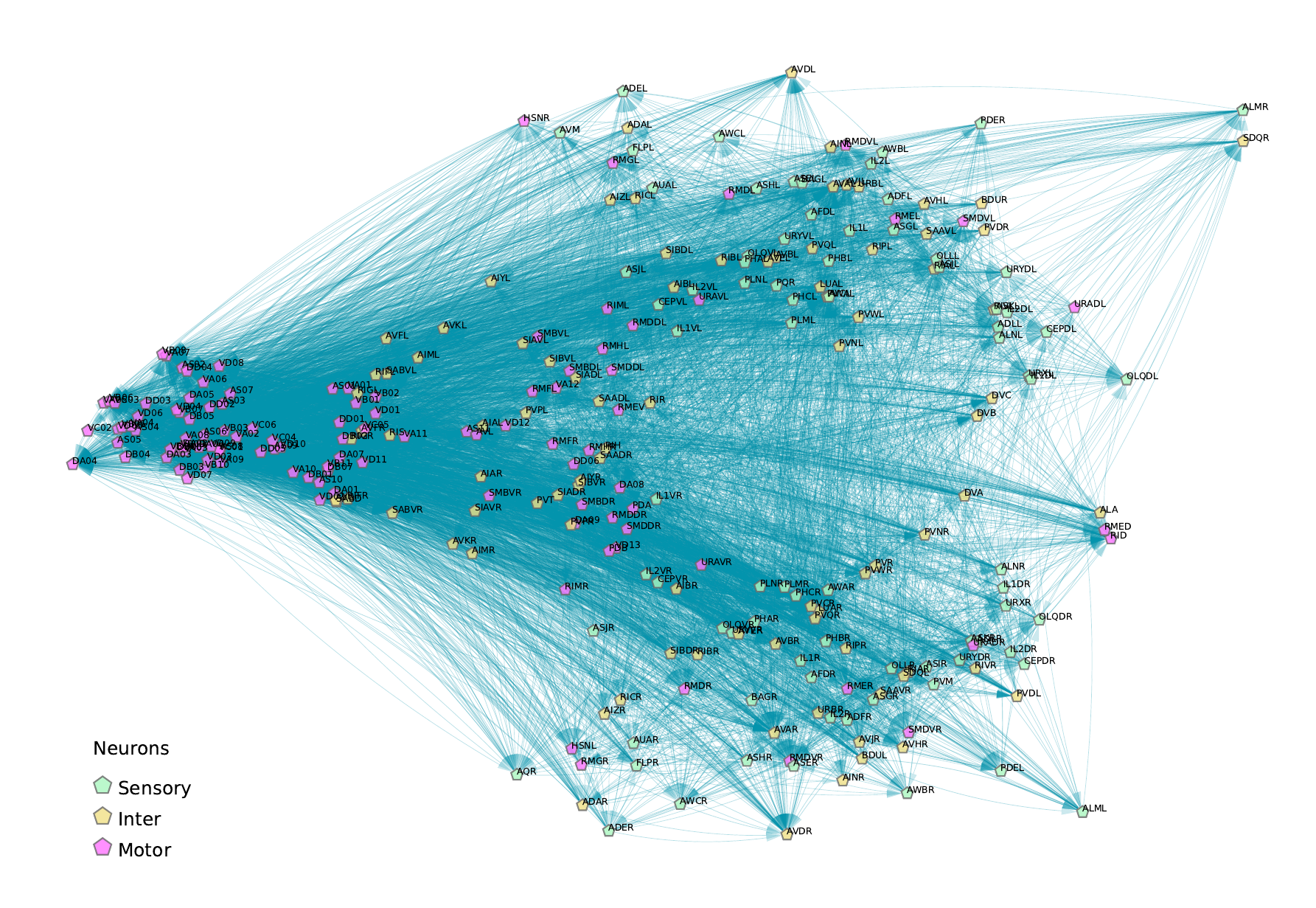}
	\caption{{\bf Purely topological functional network.} Schematic of the purely topological functional connectome $\cal F^\b$ of the \cels at a functional inverse temperature value $\b_f=4.41$. It represents only $11.68\%$ of the whole directed weighted network defined by the matrix $[\xx^{\bullet|\b}]$ whose column vectors are the neural $\b$-emittance profiles of all neurons. Neurons are positioned according to their spatial coordinates~\cite{skuhersky2022atlas}.}
	\label{fig:func-connect}
\end{figure*}

\begin{figure*}[!h]
	\centering
	\includegraphics[width=\textwidth]{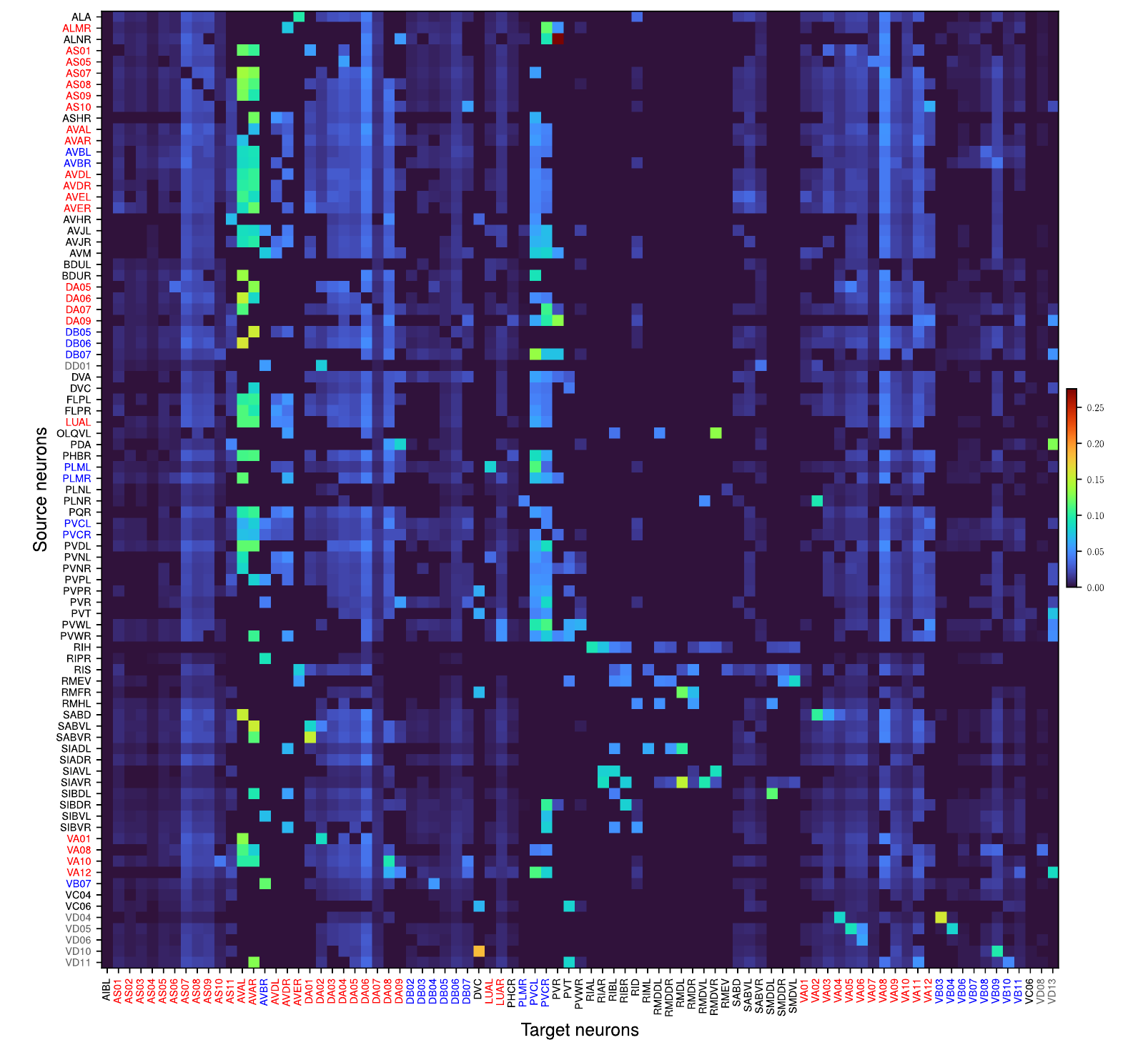}
	\caption{{\bf Purely topological functional connectivity matrix.} Representation of part of the adjacency matrix of the purely topological functional connectome $\cal F^\b$ at  $\b=4.51$. Neurons on the x-axis are the ones with the largest in-degrees in $\cal F^\b$, and on the y-axis are those with the highest out-degrees. }
	\label{fig:func-matrix}
\end{figure*}

\subsection{Classifying neurons from IC} 
 
\begin{figure*}[h]
	\centering
	\includegraphics[width=0.6\textwidth]{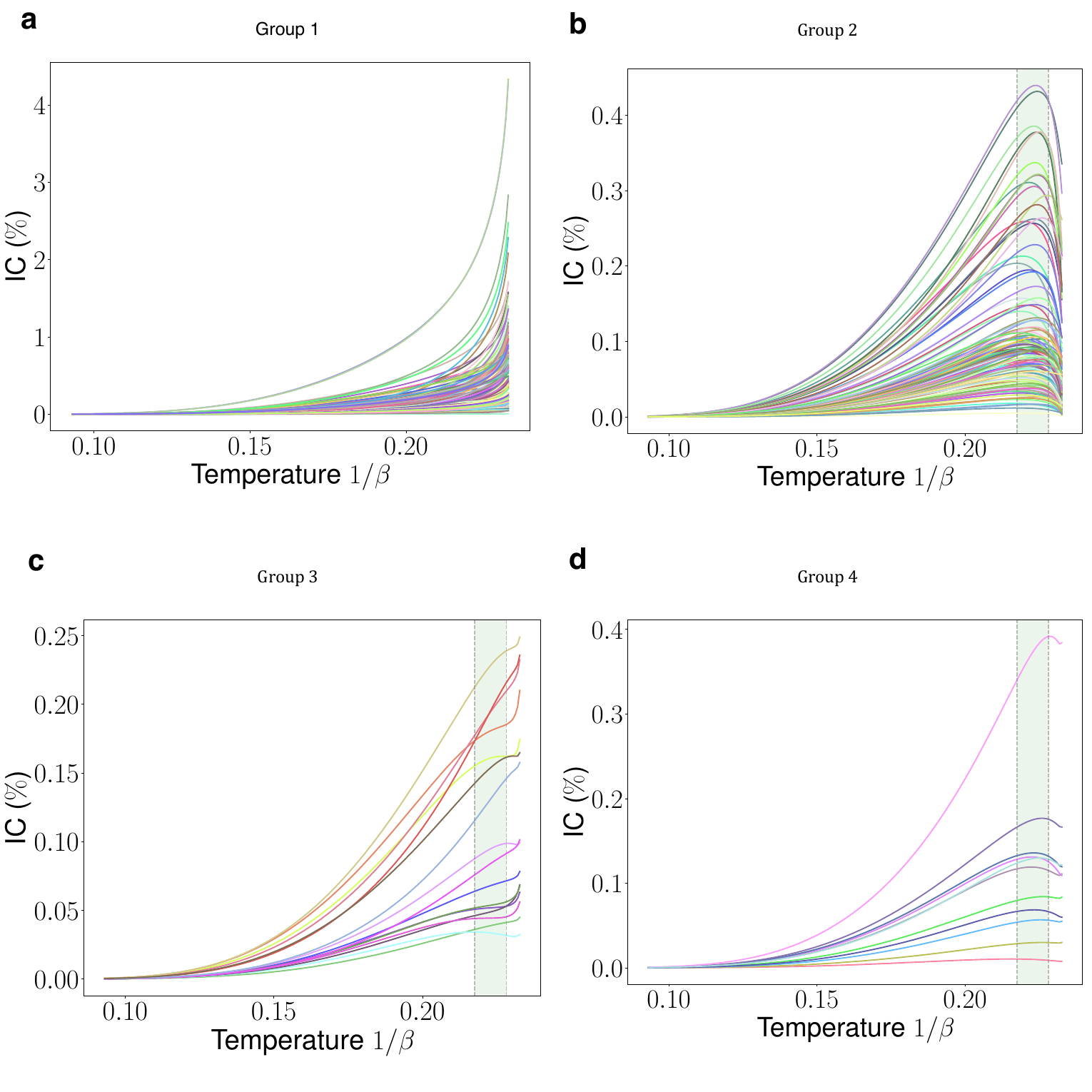}
	\caption{{\bf Integration capacity (IC).} Graphic representations of the $\IC$ of all neurons as a function of temperature. The behavior of the integration capacity function segregates the connectome into four groups of neurons, each characterized by the global shapes of the $\IC$ curves as represented by {\bf a, b, c}, and {\bf d}, respectively. The shaded area shows the {\em optimal temperature interval} within which the $\IC$ function reaches either a maximum for group 2 neurons or a near plateau for group 3 and 4.}
	\label{fig:IC}
\end{figure*}

To investigate the functional properties of the $\IC$, we analyzed its variation with respect to temperature.
We observed four main neuron groups showing different integration behaviors (see~\Cref{tab:classification} for the complete classification) schematically illustrated by Fig.~\ref{fig:IC} and characterized as follows.  

%%%%%%% Figure: IC %%%%%%%%%%%%%

\paragraph{Group 1} -  For these neurons, $\IC$ increases indefinitely with temperature, following an exponential law (Fig.~\ref{fig:IC}a). This group contains the touch receptor neurons ALMR, AVM, PLML/R, and PVM~\cite{chalfie1985neural,Goodman2006}, as well as all the neurons known to be involved in the locomotion functional circuitry~\cite{Varshney2011,white1986structure,morone2019}; namely, the command interneuron classes AVA, AVB, AVD, AVE, and PVC, and the motor neuron classes A, B, and D.

\paragraph{Group 2} - $\IC$ increases with temperature and 'saturates' at a maximum value within an inverse temperature interval $I_f$ between $\bfmin$ and $\bfmax$, before dropping to zero around the critical temperature (Fid.~\ref{fig:IC}b). This implies that redundancy among upstream pathways of these neurons becomes rare above certain path length. Cells belonging to this group include the chemosensory~\cite{Bargmann2006} neuron classes ADL, ADF, ASE, ASG, ASI, ASJ, ASK, AWA, AWB, and URX, as well as all the neuron classes  AIA, AIB, AFD, AWC, AIY, AIZ, and RIA which are known to mediate the worm's {\em thermotaxis} behavior~\cite{mori2007,kuhara2008temperature,kuhara2011,ma2012thermo}. 

\paragraph{Groups 3 and 4} - For a small number of neurons including ADAL/R, HSNR, PHAL/R, PHBR,  etc., $\IC$ increases until it reaches a near plateau within the same interval $I_f=[\bfmin, \bfmax]$, followed by an inflection point with a positive ({\em group 3}) or negative ({\em group 4}) slope, before rising again around the critical temperature (Figs.~\ref{fig:IC}c \& d). This implies that for these neurons, redundancy fluctuates among upstream pathways of above certain path length.

% TABLE : IC CLASSIFICATION
\begin{table*}[!h]
	\centering
	\caption{{\bf Classification of neurons by IC}}
	\includegraphics[width=0.6\textwidth]{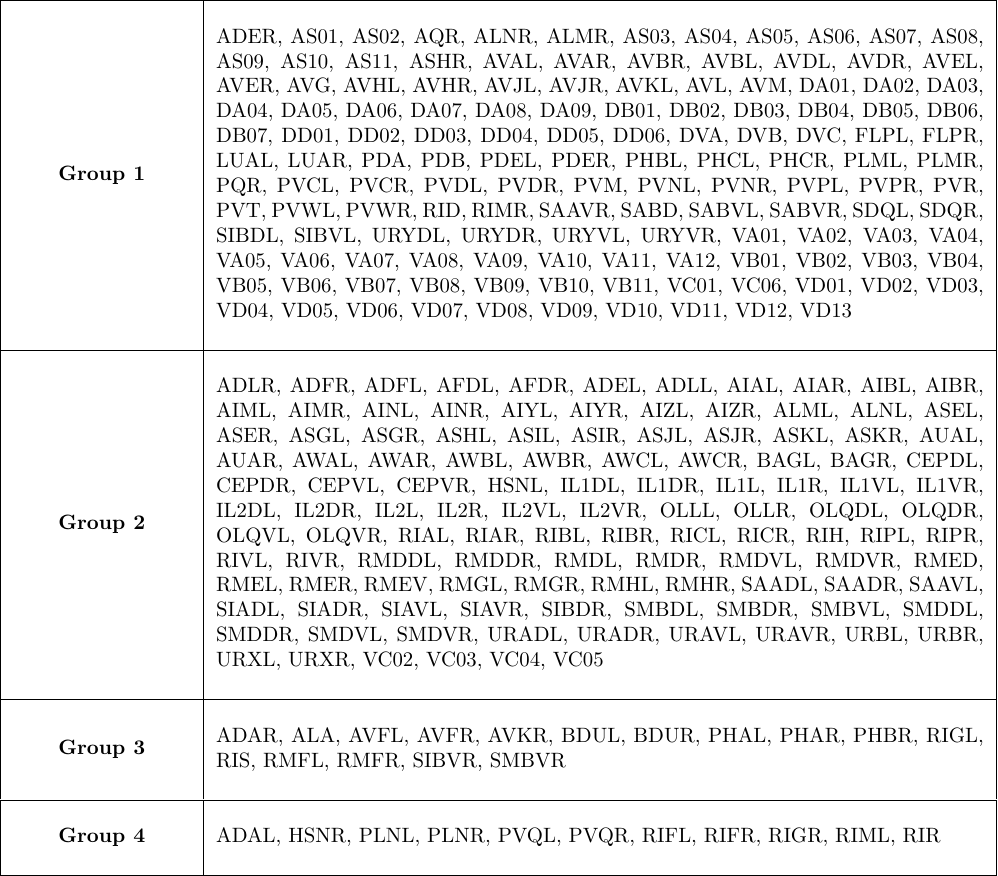}
	\label{tab:classification}
\end{table*}

\subsection{Asymmetrical integration} 
Some neuron classes show asymmetry in their integration capacity. For instance, the touch receptor neuron ALMR is in {\em group 1} while its sister cell ALML belongs to {\em group 2} (see Tab.~\ref{tab:classification}). The polymodal amphid sensory neuron class ASH, which has been shown~\cite{kaplan1993,Ghosh2017,hillard2002} to have a multisensory integration function, has its right cell ASHR in {\em group 1} and left one ASHL in {\em group 2}. While PHAL and PHAR are both in {\em group 3}, PHBL belongs to {\em group 1} and its sister cell PHBR is in {\em group 3}. This indicates that among the phasmid neuron classes PHA and PHB, which have been shown~\cite{hillard2002,jarell2012} to belong to an antagonistic functional circuit that integrates sensory responses from the amphids ASH  and ASK to mediate {\em chemo-repulsion} behavior, PHBR is structurally well positioned to have greater long-range integration function.

\section{Experimental validation}\label{sec:experimental}

\subsection{PTFC reveal entrality of the locomotory circuit}
We investigated  $\cal F^\b$ at the functional $\b_f = 4.51$ (see Supplementary Data 3 for the edgelist~\cite{MoutuouData2025}) and found that it reveals connectivity patterns that coincide with circuitries that have been reported by researchers to be the functional substrates of well studied \cels complex behaviors such as locomotion and touch-induced movement~\cite{white1986structure,chalfie1985neural}, indicating that our network model provides a theoretical basis for understanding the functional organization of a nervous system.

Indeed, $\cal F^\b$ has 8932 weighted directed edges, compared to the network $\cal K$ formed by the structural connectivity of all neurons which has 4927 weighted directed edges (see Supplementary Data 2~\cite{MoutuouData2025}). It represents $11.68\%$ of all the potential functional connections given by the adjacency matrix $ [\xx^{\bullet|\b}]$, implying that only about $12\%$ of all neural $\b$--emittance connections are PTFCs. This network is schematically represented in~\Cref{fig:func-connect}.

We observed a substantial topological discordance between $\cal F^{\b}$ and $\cal K$. For instance, the motor neuron classes AS, VA, DA, VB, and DB, have the highest in-degree centrality $\cal F^{\bf}$, with AS08, DA06, AS07, DA07, VA08, DB05, VA10, DB06, AS09, VA11 being the 10 most central neurons in terms of incoming connections, and specifically AS08, which is ranked 278 in the anatomical network (see~\Cref{tab:ranking}), receiving 219 PTFCs. Notably, none of these neurons is among the most in-connected in the structural connectome ($k^{in} = 2$ for AS08 and 7 for DA06) which has the command interneuron classes AVA, AVB, AVE, AVD, and PVC receiving the highest numbers of incoming connections. Moreover, VA08 has the highest weighted in-degree $w^{in}$ followed by DA06, and DB05 has the largest weighted out-degree $w^{out}$, while they are poorly connected anatomically. Additionally, some of these neurons are highly ranked in terms of out-degree and weighted out-degree (see~\Cref{tab:ranking}). Specifically, DA07 has the second largest number of outgoing connections ($k^{ou} = 64$), compared to its $3$ anatomical weighted connections, VB07 has out-degree rank 6 and weighted out-degree rank 7, and DB05 has out-degree rank 9 and weighted out-degree rank 1. Interestingly, PVCL and PVCR are the most functionally in- and out-connected among the command interneurons most of whom have their anatomical status downgraded in the pure functional connectome. Indeed, on the one hand, AVAL, AVAR, PVCL, and PVCR have their respective structural weighted in-degree ranks almost maintained in $\cal F^{\b}$, while their weighted out-degree ranks are not, and on the other, PVCL and PVCR have their high structural out-degree ranks almost maintained in $\cal F^{\b}$ while their weighted out-degree ranks are not.

Overall, as Fig.~\ref{fig:func-matrix} shows, the command interneurons have higher weighted degrees than the motor A and B cell types, albeit with lesser connections. Thus, both groups constitute the most central neurons of the pure functional connectome.  In other words, the ventral cord motor neuron classes AS, A, and B, are well positioned to be the target of the highest number and most uniform information flow pathways within the connectome, and the command interneuron classes AVA, AVB, AVD, AVE, and PVC send and receive the most intense flows. This theoretical result is consistent with experimental studies that have described the central role played by these neurons in locomotion~\cite{white1986structure,chalfie1985neural,morone2019}, which is the core of all \cels functions and behaviors~\cite{Bono2005,Piggott2011}. 

%  TABLE : RANKING IN FUNCTIONAL CONNECTOME
\begin{table*}[!h]
%	\scalebox{0.9}{
% \input{./tables/ranking}
%}
	\caption{{\bf Degree and weighted degree centrality.} Ranking of neurons with respect to the in- and out-degree ($k^{in}$ and $k^{out}$) and the weighted in- and out-degree ($w^{in}$ and $w^{out}$) of the pure functional connectome $\cal F^\b$ and the weighted structural network $\cal K$.}
	\includegraphics[width=0.6\textwidth]{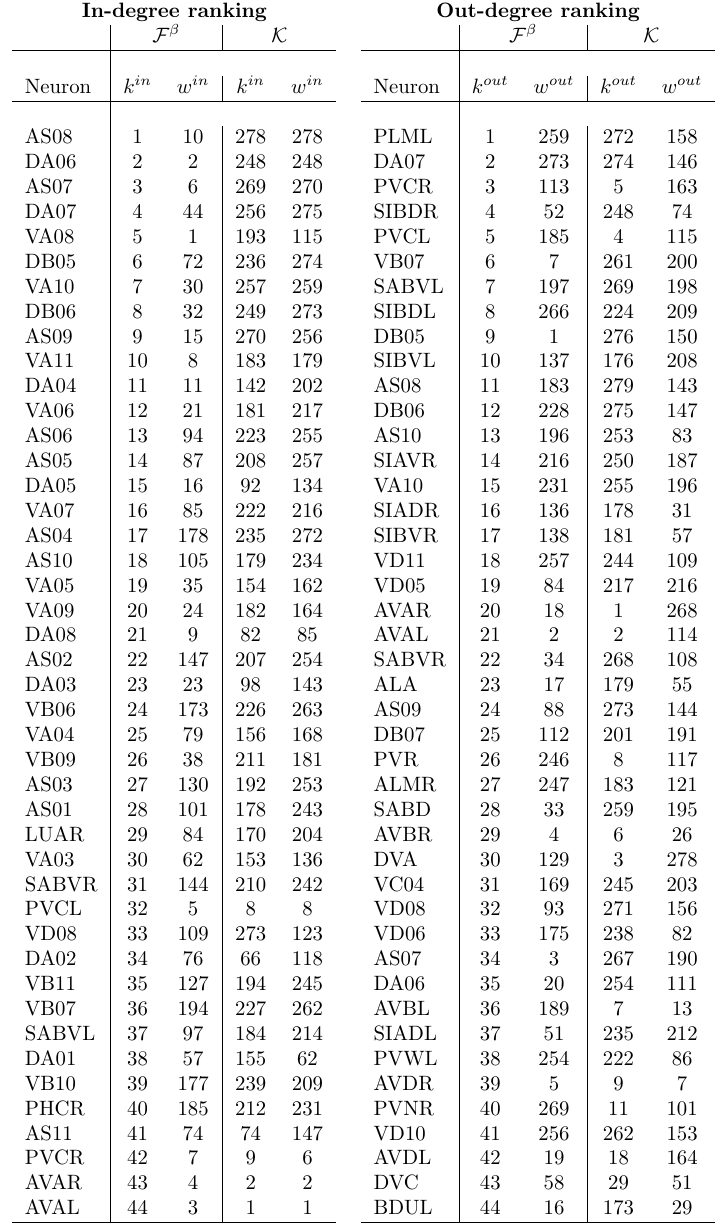}
	\label{tab:ranking}
\end{table*}

\subsection{Mechanoreceptor neurons have higher functional out-degree}

Additionally, we found that the posterior mechanoreceptor neuron~\cite{Goodman2006} PLML is the most central in number of outgoing connections in $\cal F^\b$ ($k^{out}=66$), and the anterior mechanoreceptor ALMR is among the 30 most out-connected with $k^{out} = 54$ in $\cal F^\b$ (ranked 27 in~\Cref{tab:ranking}). By comparison, both neurons have very low out-degrees of respectively 4 and 12 and low ranks of 272 and 183 in the structural network $\cal K$, and both neurons have low ranks in terms of weighted out-degree (259 and 247, respectively). More generally, many mechanoreceptor neurons~\cite{kaplan1993,Goodman2006,Wicks1995}, including ALA, ASHR, AVM, DVA, FLPL/R, OLQVL, etc., are among the most out-connected in $\cal F^\b$.

\begin{figure}[h]
	\centering
	\includegraphics[width=0.4\textwidth]{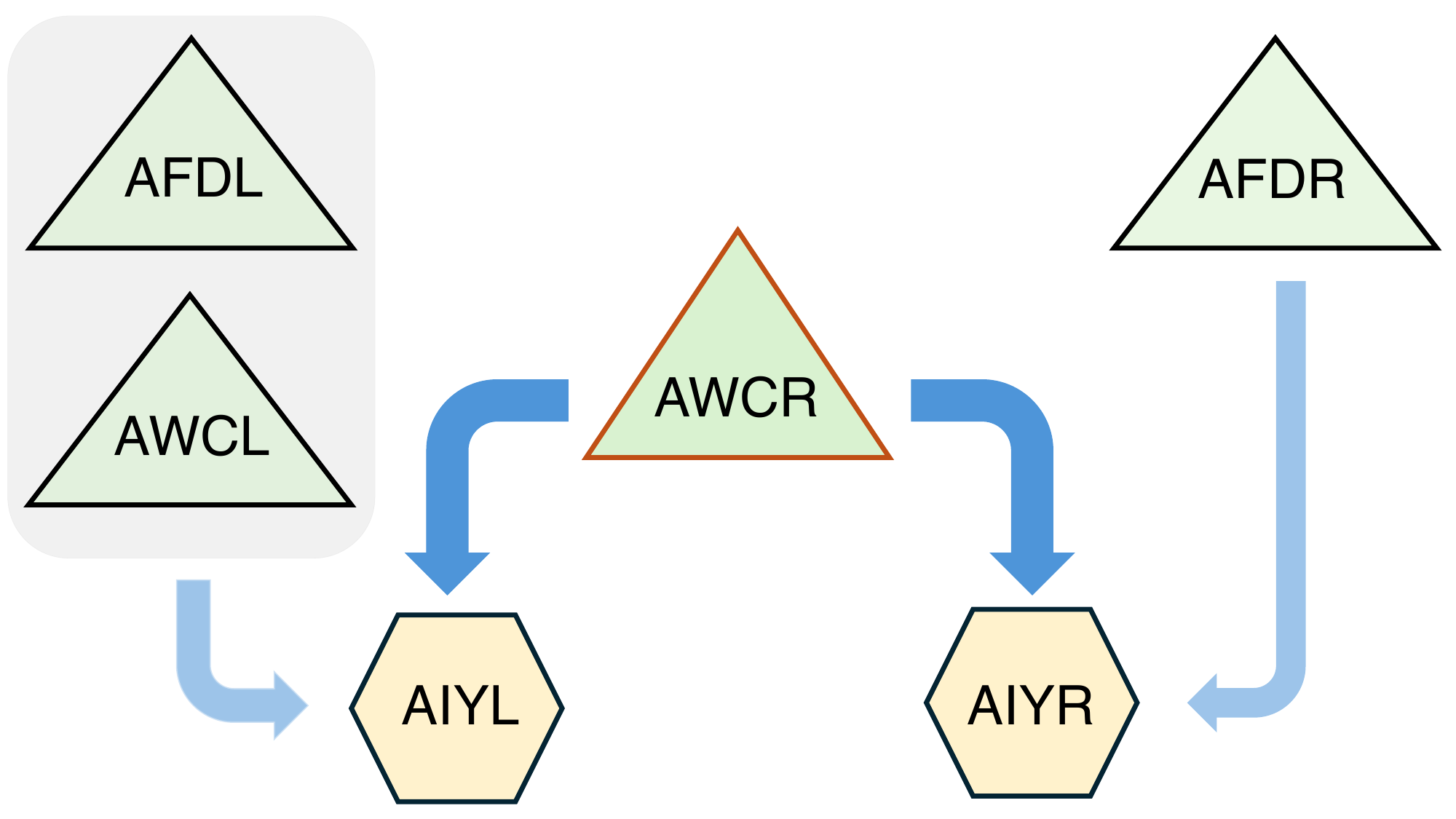}
	\caption{{\bf Asymmetric regulation of AIY integration by AFD and AWC.} The sensory neurons AFDL and AWCL only regulate the integration capacity of the left AIY (AIYL), and AWCR only regulates the right AIY. However, AWCR regulates the integration capacity of both left and right AIY.}
	\label{fig:AWCR_vs_AIY}
\end{figure}

\subsection{IC predicts response of AIY in AFD-- or AWC--ablated animals}

As a direct application of the $\IC$ measure, we theoretically reproduced an experimental study by Kano \etal~\cite{kano2023AWC} that has shown that the thermosensory neuron class AWC regulates the information processing in the AFD-AWC-AIY circuit~\cite{kuhara2008temperature,kuhara2011}, which is part of the functional circuit mediating the \cels 'thermotaxis' behavior~\cite{mori2007,ma2012thermo}.

%%%%%% AIY IC
\begin{figure}[h]
	\centering
	\includegraphics[width=0.8\textwidth]{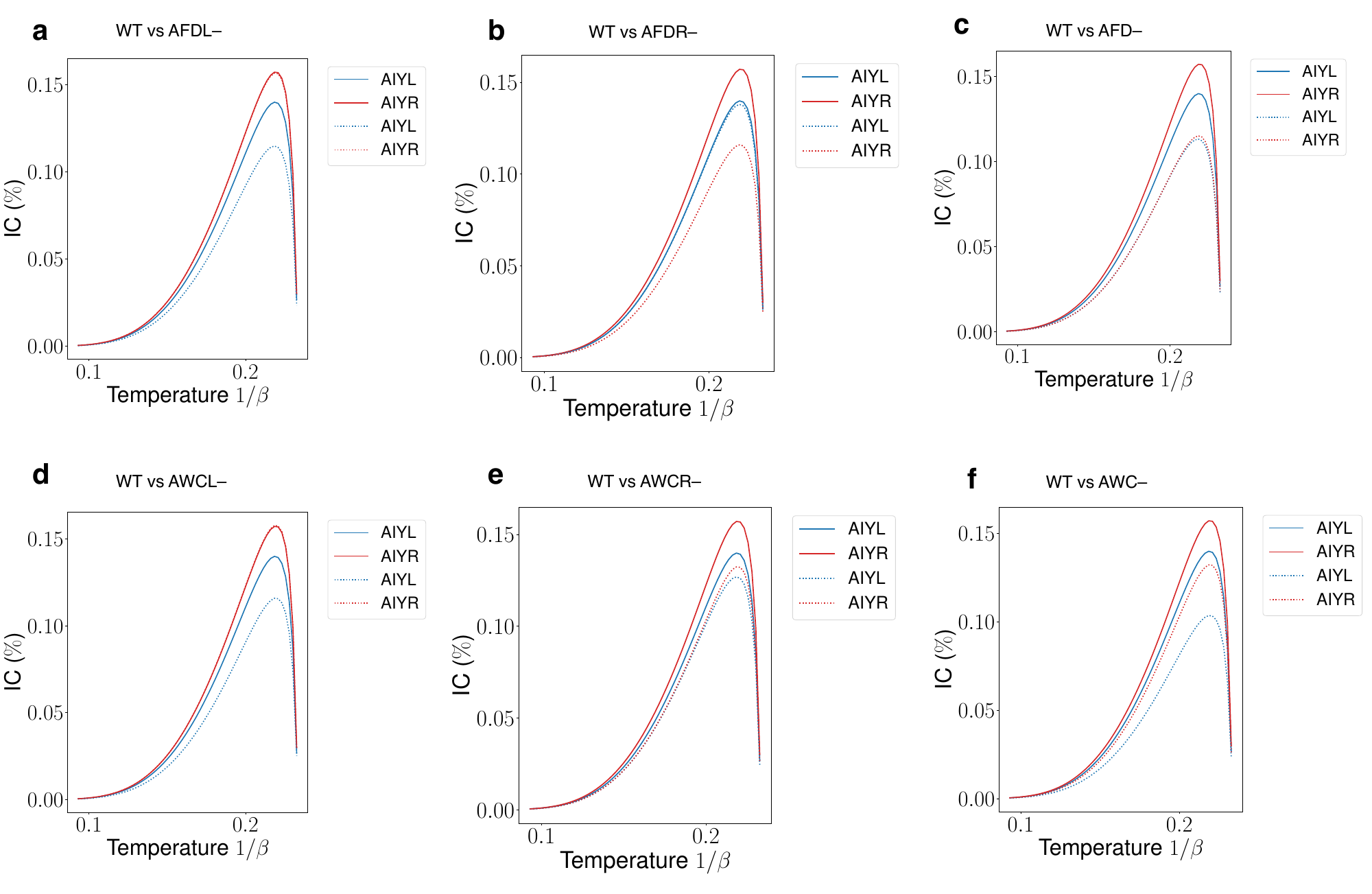}
	\caption{{\bf Change of the IC of AIY.} Comparison of the $\IC$ variation of AIYL and AIYR between the connectome of wild-type (WT) animals and the connectome of animals in which either one or both neurons in the AFD or AWC class are ablated. Continuous lines represent WT, and dashed lines represent $\IC$ when an individual neuron or a neuron class is ablated. Comparison is made using the two-sample Kolmogorov-Smirnov statistical test. In WT, AIYL and AIYR are asymmetrical in terms of their $\IC$, with AIYR's IC significantly higher than that of AIYL ($p=10^{-8}$). {\bf a}: The $\IC$ is compared between WT and ADFL-ablated: only AIYL has its $\IC$ significantly impacted (decreased). {\bf b}: WT compared to AFDR-ablated: only AIYR has $\IC$ impacted. {\bf c}: WT compared to both ADFL and AFDR ablated: both AIYL and AIYR have lower $\IC$ while their $\IC$ asymmetry is eliminated. {\bf d}: WT compared to AWCL-ablated: only  AIYL's $\IC$ decreases. {\bf e}: WT compared to AWCR-ablated: both AIYL and AIYR have lower $\IC$ and their asymmetry is significantly reduced. {\bf f}: WT compared to both AWCL and AWCR ablated: both AIYL and AIYR have lower $\IC$, while the $\IC$ asymmetry  significantly increases.}
	\label{fig:AIY} 
\end{figure}

Specifically, we study the $\IC$ variation of AIYL and AIYR within the connectome of wild-type (WT) animal and within that of $x$-ablated animal, where $x$ is either one or both of the neurons in the AFD or AWC class. We shall emphasize that in~\cite{kano2023AWC}, the authors could not investigate the individual neurons AFDL, AFDR, AWCL, AWCR, AIYL, and AIYR separately, but only considered the corresponding neuron classes. Here, our framework being based on individual neurons, we can theoretically evaluate the extent to which the removal of each individual cell impacts information flow and integration capacity of the others. Indeed, to theoretically represent the ablation of a neuron, we removed from the graph all the anatomical connections going in and out of it, an approach that allows to keep the same dimension for the resulting connectivity state vectors. We then investigated the $\IC$ of AIYL and AIYR before and after ablation by comparing their variations using a Kolmogorov--Smirnov statistical test. Results of this analysis are graphically presented in~\ref{fig:AIY}, summarized by~\Cref{fig:AWCR_vs_AIY}, and detailed as follows:
\begin{itemize}
	\item In wild-type animal connectome, there is  an $\IC$ asymmetry between AIYL and AIYR; namely, the KS-test shows that the $\IC$ of AIYR is significantly higher than that of AIYL ($p=10^{-8}$).
	\item  In AFDL- or AWCL-ablated animal connectome  (Figs.~\ref{fig:AIY}a \&~d), the $\IC$ of AIYL significantly decreases ($p<10^{-14}$), while the $\IC$ of AIYR remains the same as for wild-type connectome ($p\simeq 0.93$).
	\item Ablation of AFDR (Fig.~\ref{fig:AIY}b) significantly decreases the $\IC$ of AIYR ($p=3.10^{-23}$) but does not affect AIYL ($p=0.2$) , and ablation of AWCR (Fig.~\ref{fig:AIY}e) significantly decreases the $\IC$ of both AIYL ($p=10^{-7}$) and AIYR ($p=10^{-12}$) while significantly reducing the $\IC$ asymmetry between AIYL and AIYR  ($p=0.002$).
	\item In the connectome of animals with both AFDL and AFDR ablated (Fig.~\ref{fig:AIY}c), the $\IC$ significantly decreases in AIYL ($p=2.10^{-16}$) and AIYR ($p=8.10^{-24}$), and both AIYL and AIYR have the same $\IC$ ($p=0.12$).
	\item Ablation of both AWCL and AWCR (Fig.~\ref{fig:AIY}f) not only significantly decreases the $\IC$ of both AIYL ($p=10^{-23}$) and AIYR ($p=7.10^{-13}$) but also increases the $\IC$ asymmetry between AIYR and AIYL ($p=3.10^{-18}$).
\end{itemize}
It follows that our results not only mathematically predict the experimental observations of~\cite{kano2023AWC}, but also demonstrate that it is mostly AWCR  that is responsible for regulating information processing in the AFD-AWC-AIY circuit.

%%%% 
\section{Discussion}\label{sec:discussion}
In summary, the concept neural emittance profiles for directed networks introduced in this work captures the relative contributions of structural connectivity to the emergence of function, thereby offering a rigorous mathematical foundation for formalizing and understanding the structure-function relationship in neuroscience. More precisely, we have shown that at inverse temperatures $\b$ above the critical value $\b_c$, each neuron generates particular functional states---the pure $\b$--KMS states---that produce the emittance networks whose edge weights encode potential functional connectivity through infinite flow pathways. We addressed the important question of how these networks differ from anatomy by introducing the $\sfd$ measure, showing that for certain neurons, the emittance networks closely align with their structural wiring, while for others, they deviate significantly from their anatomical configurations. These divergent behaviors likely confer distinct anatomo-functional properties on neurons.

From the interconnectivity matrices $[\xx^{\bullet|\b}]$ formed by the pure functional states, we extracted the directed weighted networks $\cal F^\b$---the pure functional connectomes---consisting of all the neural emittances that are PTFCs; \ie, determined by the network topology of the structural connectome. In particular, at a functional inverse temperature $\b_f=4.51$, $\cal F^\b$ represents only about $12\%$ of all possible neural emittances. Extrapolating this result to undirected graphs, this connectivity density would account for approximately $24\%$ of functional connections, which is surprisingly consistent with functional connectivity thresholding in neuroimaging studies in humans and animals published in the literature~\cite{Heuvel2017}.
   
Promising engineering techniques have recently been developed for emulating synaptic plasticity and exploring the structure-function relationships~\cite{PortadelaRiva2023,Rabinowitch2024}. These observational approaches involve inserting synthetic synapses into in vivo neural circuits and investigating how the added connections interact with brain functional properties and modify animal behavior. However, these synaptic manipulations are performed under plasticity hypotheses linking local structural modifications to changes in global information flow and behavioral adaptation~\cite{Scott2022}. Moreover, while~\eqref{eq:x-v-beta} shows that adding a direct connection from neuron $u$ to $v$ strengthens the neural emittance of $u$ to $v$, we observed that structural connections do not necessarily underlie PTFCs, and conversely, PTFCs do not always align with structural connections. This implies that not all insertions of synthetic synapses are likely to generate functional connections. Therefore, although synaptic plasticity was not the focus of this work, our framework can guide the identification of changes in neuronal structural interconnectivities that lead to functional network reconfigurations, offering a practical atlas of relevant neuron pairs for more efficient synaptic engineering.

On the other hand, experimental research in \cels has identified functional interactions among neurons that differ from predictions based on anatomical information flow pathways, with evidence showing that extrasynaptic transmission contributes to these differences~\cite{randi2023celegans}. This indicates that direct connections via extracellular neurotransmitters~\cite{bentley2016multilayer} dynamically and selectively add to the original graph, altering its computational and functional properties by establishing new functional connections. This property can be simulated by extracting from the connectivity matrix $[\xx^{\bullet|\b}]$ neural emittance connections that (i) are not mapped to existing structural connections and (ii) do not belong to the pure functional connectome $\cal F^\b$. New direct edges representing virtual extrasynaptic connections are then added between the corresponding neuron pairs to investigate their impact on the pure functional connectome. Indeed, performing such a simulation on RID revealed the emergence of new PTFCs that correspond with the extrasynaptic signaling-based functional connectivity identified by Randi \etal~\cite{randi2023celegans}. 

This approach not only describes how network computation and the functional properties of the nervous system are modified by dynamic extrasynaptic interactions, but also rigorously demonstrates that the physical synaptic connectome does not completely determine function. Further simulations using the neuropeptidergic connectome data published by Ripoll-S\`anchez \etal~\cite{ripoll2022} could provide additional insights into how extracellular connectivity influences the \cels pure functional connectome. Additionally, extending our framework to complex systems with higher order interactions and multilayer networks~\cite{kivela2014,moutuou2023} could enable the analysis of the emittance networks generated by neuronal assemblies and to investigate the intricate relationship between the wired synaptic circuitry and the wireless connectomes such as the one defined by the neuropeptidergic signaling.

Furthermore, our results from the integration capacity analysis and simulations of individual cell ablation within the thermotaxis neuronal circuit theoretically explain the in vivo studies of Kano \etal~\cite{kano2023AWC} and reveal the functional asymmetry of the thermosensory neuron class AWC, which regulates information processing in the AFD-AWC-AIY sub-circuit, as well as the interneuron class AIY, paving the way for further investigations into the functional differences among neurons of the same class and how the activity of one neuron influences another.

Finally, although we used the \cels connectome to illustrate our work, the formalism developed here is broad and can be applied to general directed complex networks to understand their functional properties.

\section{Materials and Methods}\label{sec:methods}

\paragraph{The \cels somatic connectome.} We have merged two datasets of the somatic connectome of the hermaphrodite worm \cels that are publicly available on the WormAtlas~\cite{wormatlas2023}: the first dataset, {\em Data1}, is from Varshney \etal~\cite{Varshney2011}, and the second one, {\em Data2}, is from the recently published serial electron microscopy reconstructions by Cook \etal~\cite{cook2019connectome}. Specifically, we have kept all synaptic connections from {\em Data1} (279 connected neurons and 8171 chemical and electrical synapses), to which we have added all synaptic connections in {\em Data2}  that were not originally in {\em Data1} (3900 additional connections). The resulting connectome consists of 280 connected neurons and 12071 synaptic connections. Note that in the old version ({\em Data1}), VC06 had no connections, while it is connected in the revised connectome.

\medskip

\paragraph{Computing the structure-function divergence and the mean total receptance.} To calculate $\sfd(v,\b)$ for a neuron from Eq.~\ref{eq:sfd}, we excluded self-loops by using the connectivity distribution $\bar{\kk}^v$ instead of $\kk^v$, where $\bar{\kk}^v_v=0$ and for $u\neq v$, $\bar{\kk}^v_u=\kk^v_u/ \sum_{w\neq v}\kk^v_w$ if $\kk^v_u>0$. 

\medskip 

\paragraph{Statistical significance of neural emittance.} In order to know the extent to which a given neural $\b$--emittance is statistically significant, we used a bootstrapping technique consisting of generating 5000 random directed graphs with the same degree sequence as the graph $G$ representing the connectome, and computing the neural emittance of the same neuron in each generated graph at the same inverse temperature $\b$. The neural emittance is then statistically significant if it occurs with the same or larger value in less than $5\%$ of the time ($p<0.05$).

\medskip 

\paragraph{Data and code availability. } The code written for this work and the data generated are available in the public domain at \href{https://github.com/elkMm/KMSnet}{https://github.com/elkMm/KMSnet}.

\medskip

\subsection*{Acknowledgment} This work was supported by the Natural Sciences and Engineering Research Council of Canada through the CRC grant NC0981.

%
% \bibliographystyle{ieeetr}
% \bibliography{../../bib/biblio}

\end{document}